\documentclass[icps3]{svjour}
\usepackage{epsfig}

\begin{document}
\title{Coulomb drag in phase-coherent mesoscopic structures}

\subtitle{--- Numerical study of disordered 1D--wires}

%\titlerunning{Coulomb drag in phase-coherent mesoscopic structures}
\titlerunning{Proc. 25th Int. Conf. Phys. Semicond., Osaka 2000 (eds. N. Miura and T. Ando) {\it Springer Proc. in Phys. {\bf 87} (2001)}}

\author{Niels Asger Mortensen\inst{1,2} \and Karsten Flensberg\inst{2}
  \and Antti-Pekka Jauho\inst{1}}

\authorrunning{N.A. Mortensen, K. Flensberg, and A.-P. Jauho}

\institute{Mikroelektronik Centret, Technical University of Denmark,
  \O rsteds Plads bld. 345 east, DK-2800 Kgs. Lyngby, Denmark \and \O
  rsted Laboratory, Niels Bohr Institute, University of Copenhagen,
  Universitetsparken 5, DK-2100 Copenhagen \O, Denmark}

\setcounter{page}{1347}
\maketitle

\begin{abstract}
  We study Coulomb drag between two parallel disordered mesoscopic
  1D--wires. By numerical ensemble averaging we calculate the
  statistical properties of the transconductance $G_{21}$ including
  its distribution. For wires with mutually uncorrelated disorder
  potentials we find that the mean value is finite, but with
  comparable fluctuations so that sign-reversal is possible. For
  identical disorder potentials the mean value and the fluctuations
  are enhanced compared to the case of uncorrelated disorder.
\end{abstract}

\section{Introduction}\label{introduction}

Current flow in a conductor can through a Coulomb mediated drag-force
accelerate charge-carriers in a nearby conductor, thus inducing a
drag-current. The effect is active whenever the distance between the
two conductors is of the same order as the distance between the
charge-carriers -- otherwise it is suppressed by screening. In the
past years Coulomb drag in extended 2D-systems has been studied
extensively~\cite{rojo} and very recently the study of fluctuations of
the Coulomb drag was initiated by Narozhny and Aleiner~\cite{narozhny}
who found that the fluctuations will be pronounced for temperatures
smaller than the Thouless energy. We study how drag between disordered
mesoscopic 1D--wires~\cite{morta,mortb} give rise to these new
interesting phenomena such as a large fluctuations and sign reversal
of the drag current.

\section{Formalism}\label{formalism}

\begin{figure}
\begin{center}
\epsfig{file=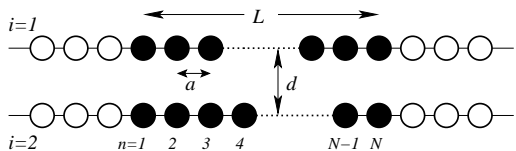, width=0.8\columnwidth,clip}
\end{center}
\caption{Coulomb coupled 1D--wires where $\bullet$ denote the lattice points of the wires and $\circ$ denote those belonging to the ideal leads.}
\label{fig:sample}
\end{figure}

Consider two 1D--wires of length $L$ (shorter than the phase-breaking
length) parallel to each other with a separation $d$, see
Fig.~\ref{fig:sample}. Writing the Laplacian with the help of finite
differences the Hamiltonian of the uncoupled wires is mapped onto a
tight-binding model~\cite{datta}

\begin{equation}
\big\{H_i\big\}_{nn'}=[2t+U_i(n)]\delta_{nn'} 
-t\delta_{n,n'\pm 1}\,,\, t=\frac{\hbar^2}{2ma^2},
\end{equation}
where $i=1,2$ label the two wires and $n,n'=1,2,3,\ldots N$ the
lattice points. The conducting properties can be obtained from the
retarded Green functions of the isolated wires which can be written
as $N\times N$ matrices: ${\cal G}_i=(\varepsilon_{F} - H_i
-\Sigma_L^i- \Sigma_R^i)^{-1}$, where $\Sigma_p^i$ is the retarded
self-energy describing coupling to the lead $p=L,R$.

Using Kubo formalism we calculate the Coulomb drag to second order in
the interaction $U_{12}$ between the mesoscopic 1D-wires which we
assume to be otherwise non-interacting~\cite{kamenev,flensberg}. For
$kT\ll\varepsilon_F$ the dc transconductance $G_{21}=\partial
I_2/\partial V_1$ becomes~\cite{morta,mortb}
\begin{subequations}\label{G21}
\begin{equation}
G_{21}=\frac{e^2}{h}\left(kT\right)^2 \frac{t^2}{3} {\rm
Tr}\big[ U_{12}\, M_1\,U_{12}\,M_2\big],
\end{equation}
where $U_{12}$ is an $N\times N$ coupling matrix representing the
interwire Coulomb interaction and $M_i$ is an $N\times N$ matrix

\begin{equation}
M_i={\rm Re}\big\{ A_i^T\otimes \left[A_i\Lambda A_i\right]\big\},\;A_i=i\big[ {\cal G}_i-{\cal G}_i^\dagger\big].
\end{equation}
\end{subequations}
Here, $\Lambda_{nn'}=\pm\delta_{n,n'\pm 1}/(N-1)$ and $\big\{X\otimes
Y\big\}_{nn'}=X_{nn'}Y_{nn'}$. The Landauer conductance
$G_{ii}=\partial I_i/\partial V_i$ of the individual wires can be
expressed in a similar form~\cite{datta}

\begin{equation}
G_{ii}=\frac{2e^2}{h}{\rm Tr}\big[\Gamma_{L}^i\, {\cal G}_i
\,\Gamma_{R}^i\, {\cal G}_i^\dagger \big]\,,
\, \Gamma_{p}^i=i\big[\Sigma_{p}^i -
  \big\{\Sigma_{p}^i\big\}^\dagger\big].
\end{equation}

\section{Ensemble averaging}

The statistical properties of drag can be analyzed by generating an
ensemble of different disorder configurations and using
Eq.~(\ref{G21}) to calculate the drag. For the disorder we use the
Anderson model with diagonal disorder~\cite{anderson} where the
transport mean free path $\ell$ can be related to the disorder
strength $W$ by $\ell=a 12(4t \varepsilon_F - \varepsilon_F^2)/W^2$.
We consider two cases: {\it i)} both wires being disordered, but with
$U_1$ and $U_2$ fully uncorrelated and {\it ii)} both wires being
disordered and fully correlated~\cite{gornyi}, {\it i.e.}  $U_1=U_2$.
For weak disorder a diagrammatic perturbation expansion for the
fluctuations $\delta G_{21}=G_{21}- \left<G_{21}\right>$ gives
$\big<\big[\delta G_{21}\big]^2\big>^{1/2} \propto
1/k_F\ell$~\cite{morta,mortb} and it can also be argued that
$\big<\big[\delta G_{21}\big]^2\big>_{\rm c} = 2\times
\big<\big[\delta G_{21}\big]^2\big>_{\rm uc}$~\cite{mortb}. Both
predictions are valid to lowest order in $1/k_F\ell$.

\begin{figure*}
\begin{center}
\epsfig{file=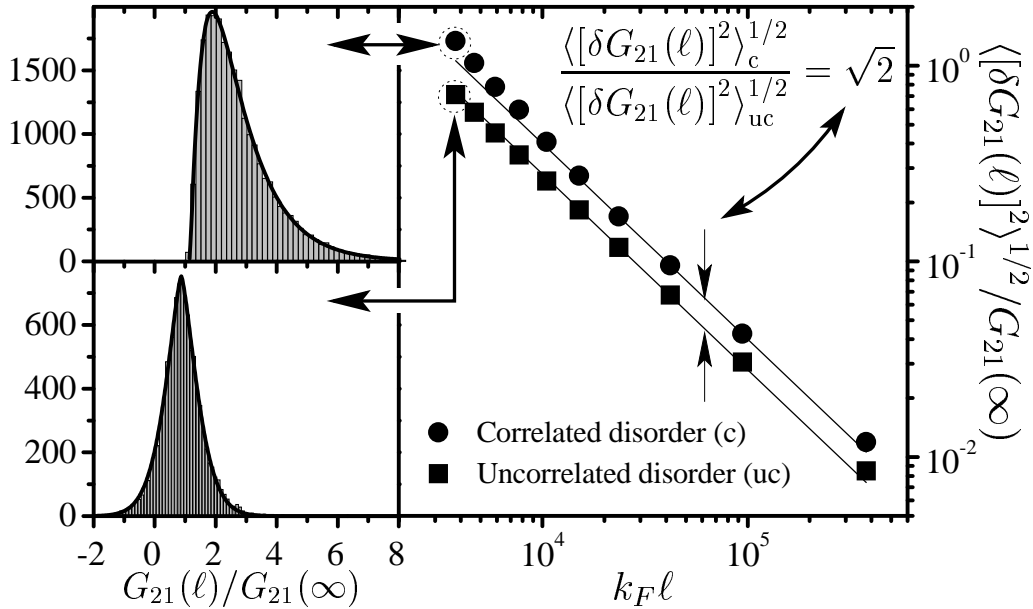, width=1.6\columnwidth,clip}
\end{center}
\caption{Right panel: fluctuations $\big<\delta G_{21}^2(\ell)\big>^{1/2}$ normalized by the ballistic result $G_{21}(\infty)$ as a function of $k_F\ell$ (see text for system parameters). Left panels: histograms (based on $\sim 10^4$ random disorder configurations) for $G_{21}(\ell)$ in the case of $k_F\ell =(\pi/3)\times 3600$. }
\label{fig:histograms}
\end{figure*}

\section{Results}

We consider quarter-filled bands ($\varepsilon_F=t$) and wires with
$N=100$ lattice points so that $k_FL =(\pi/3)\times 100$. The
separation is $k_F d= 1$ and for simplicity we assume an unscreened
coupling of the form $\big\{U_{12}\big\}_{nn'}=e^2\big/\big(4\pi
\epsilon_0\epsilon_r [(n-n')^2 a^2 +d^2]^{1/2}\big)$.

We study the de-localized regime $\ell\gg L$ where we as
expected~\cite{abri81,beenakker} find that the fluctuations $\delta G_{ii}$ of
the Landauer conductance $G_{ii}$ are vanishing and
$\big<G_{ii}\big>\simeq 2e^2/h$. However, for the transconductance
$G_{21}$ even weak disorder can have a large effect. The lower left
panel of Fig.~\ref{fig:histograms} shows a typical histogram of
$G_{21}(\ell)/G_{21}(\infty)$ (where $G_{21}(\infty)$ is the result in
the ballistic regime, $U_1=U_2=0$) for $\ell =36 L$. Depending on the
disorder configuration $G_{21}(\ell)$ can be either higher or lower
than in the ballistic regime. The enhancement occurring for certain
disorder configurations can be understood physically as follows. The
lack of translational invariance allows forward scattering
(transferred momentum $q\simeq0$), which normally has little effect,
to cause transitions between scattering states with opposite
directions, thus contributing to the drag. The variance is of the same
order as the mean value so that sign reversal for some disorder
realizations is possible. The latter is represented by the negative
tail in the histogram.

For the same system parameters but now with identical (correlated)
disorder potentials we get a very different distribution as seen in
the upper left panel of Fig.~\ref{fig:histograms}. As
predicted~\cite{gornyi} the mean value is enhanced compared to
uncorrelated disorder and also the fluctuations are enhanced. In the
right panel of Fig.~\ref{fig:histograms} we show the dependence of the
fluctuations on the mean free path $k_F\ell$ which has the expected
$1/k_F\ell$ behavior. Comparing the two disorder situations we find
numerical support for the predicted relative strength of
$\sqrt{2}$~\cite{mortb}.

\section{Conclusion}

We have numerically studied drag of disordered mesoscopic 1D-wires in
the de-localized regime $\ell\gg L$. Our results illustrate how the
statistics of the transconductance depend strongly on disorder and we
find that even weak disorder can give rise to fluctuations of the same
order of magnitude as the transconductance for the ballistic case.
This implies that the direction of drag depends on the disorder
configuration and that for a given system the sign of the drag current
will be arbitrary.  Our results also confirm the for 2D extended
systems recently predicted enhancement of the mean value for
correlated disorder compared to uncorrelated disorder. In addition we
have also found a corresponding enhancement of the fluctuations by a
factor of $\sqrt{2}$ compared to uncorrelated disorder.

\vspace{3mm} We acknowledge C.~W.~J. Beenakker and M. Brandbyge for
useful discussions.


\begin{thebibliography}{}
  
\bibitem{rojo} For a review see {\it e.g.} A.~G. Rojo, J. Phys.:
  Condens Matter \textbf{11} (1999) R31.
  
\bibitem{narozhny} Y.~V. Narozhny and I.~L. Aleiner, Phys. Rev. Lett.
  \textbf{84} (2000) 5383.
  
\bibitem{kamenev} A. Kamenev and Y. Oreg, Phys. Rev. B \textbf{52}
  (1995) 7516.
  
\bibitem{flensberg} K. Flensberg, B.~Y.-K. Hu, A.-P. Jauho, and J.~M.
  Kinaret, Phys. Rev. B \textbf{52} (1995) 14761.
  
\bibitem{morta} N.~A. Mortensen, K. Flensberg, and A.-P. Jauho,
  cond-mat/0007046.
  
\bibitem{mortb} N.~A. Mortensen, K. Flensberg, and A.-P. to be
  published.
  
\bibitem{datta} S. Datta, \textit{Electronic Transport in Mesoscopic
    Systems} (Cambridge University Press, Cambridge, 1995).
    
\bibitem{anderson} P.~W. Anderson, Phys. Rev. \textbf{109} (1958)
  1492.
  
\bibitem{gornyi}I.~V. Gornyi, A.~G. Yashenkin, and D.~V. Khveshchenko,
  Phys.  Rev. Lett. {\bf 83} (1999) 152.
  
\bibitem{abri81} A.~A. Abrikosov, Solid State Commun. \textbf{37}
  (1981) 997.
  
\bibitem{beenakker} C.~W.~J. Beenakker, Rev. Mod. Phys. \textbf{69}
  (1997) 731.

\end{thebibliography}
\end{document}